\documentclass{appolb}
\usepackage{graphicx}
\usepackage{amsmath,bm}

\begin{document}
\title{Skyrme Functional with Tensor Terms from \textit{ab initio} Calculations: Results for the Spin-Orbit Splittings%
\thanks{Presented at XXV NUCLEAR PHYSICS WORKSHOP, Structure and dynamics of atomic nuclei, Kazimierz Dolny, Poland, 25-30 September 2018}%
}
\author{Shihang Shen, Gianluca Col\`o, Xavier Roca-Maza
\address{Dipartimento di Fisica, Universit\`a degli Studi di Milano, Via Celoria 16, I-20133 Milano,  Italy}
\address{INFN, Sezione di Milano, Via Celoria 16, I-20133 Milano, Italy}
\\
}
\maketitle
\begin{abstract}
A new Skyrme functional including tensor terms is presented.
The tensor terms have been determined by fitting the results of relativistic Brueckner-Hartree-Fock (RBHF) studies on neutron-proton drops.
Unlike all previous studies, where the tensor terms were usually determined by fitting to experimental data of single-particle levels, the pseudodata calculated by RBHF does not contain beyond mean-field effect such as the particle-vibration coupling and therefore can provide information of the tensor term without ambiguities.
The obtained new functional, named SAMi-T, can describe well ground state properties such as binding energies, radii, spin-orbit splittings, and at the same time the excited state properties such as those of the Giant Monopole Resonance (GMR), Giant Dipole Resonance (GDR), Gamow-Teller Resonance (GTR), and Spin-Dipole Resonance (SDR).
\end{abstract}
  
\section{Introduction}

Even though in the past decades the nuclear \textit{ab initio} calculations have made extensive progresses, they are still difficult to be applied to the whole range of nuclear chart with high accuracy.
The nuclear energy density functional theory (DFT) remains the only tool to describe accurately at the same time the ground state and excited state properties along the whole nuclear chart in a unified way \cite{Bender2003,Meng2016,Roca-Maza2018a}.
On the other hand, many open questions still exist in nuclear DFT.
One of them, which is also the main focus of this contribution, is the tensor term \cite{Sagawa2014}.

In the early times, the tensor term in nuclear density functional was deemed not important and often discarded \cite{Vautherin1972,Walecka1974,Decharge1980}.
It attracted attention when the specific evolution pattern of single-particle (s.p.) levels in the Sn isotopes and $N=82$ isotones was discovered \cite{Schiffer2004} and could be explained by the effect of tensor term \cite{Otsuka2005}.
In contrast, none of the functionals without tensor term could reproduce this pattern \cite{Brown2006,Colo2007,Brink2007,Lesinski2007,LongWH2008}.
However, it is difficult to single out the effect of tensor term by just looking at the experimental s.p. levels as beyond-mean-field effect, e.g., particle-vibration coupling (PVC), has also strong influence on these levels \cite{Colo1994,NiuYF2012,Litvinova2011,Afanasjev2015}.

In this context, we propose to determine the tensor term not by fitting to experimental s.p. levels, but by fitting to pseudodata from \textit{ab initio} calculations.
In a recent work, an ideal system, the neutron drops, has been studied by relativistic Brueckner-Hartree-Fock (RBHF) theory \cite{Shen2016,Shen2017,Shen2018a} using the Bonn interactions and a clear signature of tensor term has been illustrated in the evolution of spin-orbit (SO) splittings \cite{Shen2018,Shen2018b}.
Along this direction, we further studied the neutron-proton drops with RBHF theory, which is also an ideal system confined in an external field without consideration of center-of-mass correction nor Coulomb interaction \cite{Shen2018c}.
Then we developed a new Skyrme functional with tensor term by fitting to the evolution of SO splittings of the neutron-proton drops from RBHF \cite{Shen2018c}.
In this way, we can extract the information of tensor term without ambiguity, as there is no beyond-mean-field effect such as PVC in the RBHF calculation.

We will briefly describe the tensor term in Skyrme functional in Sec.~\ref{sec:formula}.
Some results of the spin-orbit splittings are presented in Sec.~\ref{sec:result}.
Finally, we give the conclusions in Sec.~\ref{sec:sum}.

\section{Tensor Term in Skyrme Functional}\label{sec:formula}

The Skyrme effective interaction with two-body tensor term is written in the standard form as~\cite{Vautherin1972,Stancu1977}.
{\small
\begin{align}
V(\mathbf{r}_1,\mathbf{r}_2) &= t_0(1+x_0P_\sigma) \delta(\mathbf{r}) + \frac{1}{2}t_1
(1+x_1P_\sigma) \left[ {\mathbf{P}'}^2\delta(\mathbf{r}) + \delta(\mathbf{r}) \mathbf{P}^2 \right]
+ t_2(1+x_2P_\sigma) \mathbf{P}' \cdot \delta(\mathbf{r}) \mathbf{P},  \notag \\
&~~~+ \frac{1}{6}t_3 (1+x_3P_\sigma) \rho^\gamma(\mathbf{R}) \delta(\mathbf{r})
+ iW_0(\bm{\sigma}_1+\bm{\sigma}_2) \cdot \left[ \mathbf{P}'\times \delta(\mathbf{r}) \mathbf{P} \right] + V_T(\mathbf{r}_1,\mathbf{r}_2), \label{eq:vskyrme} \\
V_T(\mathbf{r}_1,\mathbf{r}_2) &= \frac{T}{2} \left\{\left[(\bm{\sigma}_1\cdot\mathbf{P}')
(\bm{\sigma}_2\cdot\mathbf{P}')-\frac{1}{3}(\bm{\sigma}_1\cdot\bm{\sigma}_2)
{\mathbf{P}'}^{2}\right]\delta(\mathbf{r})
+\delta(\mathbf{r})\left[(\bm{\sigma}_1\cdot\mathbf{P})
(\bm{\sigma}_2\cdot\mathbf{P})-\frac{1}{3}(\bm{\sigma}_1\cdot\bm{\sigma}_2)
{\mathbf{P}}^{2}\right]\right\} \notag \\
&+U\left\{(\bm{\sigma}_1\cdot\mathbf{P}')\delta(\mathbf{r})
(\bm{\sigma}_2\cdot\mathbf{P})-\frac{1}{3}(\bm{\sigma}_1\cdot\bm{\sigma}_2)
\left[\mathbf{P}'\cdot\delta(\mathbf{r})\mathbf{P}\right]\right\},
\end{align}
}
where $\mathbf{r} = \mathbf{r}_1 - \mathbf{r}_2, \mathbf{R} = \frac{1}{2}(\mathbf{r}_1 + \mathbf{r}_2), \mathbf{P} = \frac{1}{2i}(\nabla_1-\nabla_2)$, $\mathbf{P}'$ is the hermitian conjugate of $\mathbf{P}$ acting on the left.
The spin-exchange operator reads $P_\sigma = \frac{1}{2}(1+\bm{\sigma}_1\cdot\bm{\sigma}_2)$, and $\rho$ is the total nucleon density.

The Hartree-Fock equations for each s.p. level can be obtained by the variational method with respect to the HF total energy as,
\begin{equation}\label{eq:}
  \left[ -\frac{\hbar^2}{2M}\nabla^2 + U_q(\mathbf{r}) \right] \psi_k(\mathbf{r}) = e_k \psi_k(\mathbf{r}),
\end{equation}
where $e_k$ is the single-particle energy, $\psi_k$ is the corresponding wave function, 
and $q = 0(1)$ labels neutrons (protons).
The single-particle potential $U_q(\mathbf{r})$ is a sum of central, Coulomb and spin-orbit terms,
\begin{equation}\label{eq:}
  U_q(\mathbf{r}) = U_{q}^{\rm(c)}(\mathbf{r}) + \delta_{q,1}U_{C}(\mathbf{r}) + \mathbf{U}_{q}^{\rm(s.o.)}(\mathbf{r}) \cdot (-i)(\nabla\times\bf{\sigma}).
\end{equation}
For the calculation of proton-neutron drops, we add an external harmonic oscillator field.
The spin-orbit term reads \cite{Stancu1977,Sagawa2014}
\begin{equation}\label{eq:Uso}
  \mathbf{U}_q^{\rm(s.o.)}(\mathbf{r}) = \frac{1}{2} \left[ W_0\nabla\rho + W_0'\nabla\rho_q \right]
  + \left[ \alpha \mathbf{J}_q + \beta\mathbf{J}_{1-q} \right],
\end{equation}
where $\mathbf{J}(\mathbf{r})$ the spin-orbit density. 
Starting from Eq.~(\ref{eq:vskyrme}) one would derive $W_0' = W_0$.
We adopt a more general form and assume that $W_0'$ can be defined and fitted independently as in the case of SAMi functional \cite{Roca-Maza2012} among the others.

The parameters $\alpha$ and $\beta$ in Eq.~(\ref{eq:Uso}) include contributions from 
the (exchange part of the) central term and from the tensor term,
\begin{equation}\label{eq:alpha-beta}
  \alpha = \alpha_c + \alpha_T,\quad \beta = \beta_c + \beta_T,
\end{equation}
where
\begin{subequations}\label{eq:}\begin{align}
  \alpha_c &= \frac{1}{8}(t_1-t_2) - \frac{1}{8}(t_1x_1+t_2x_2),\quad
  \alpha_T = \frac{5}{12}U,\\
  \beta_c &= -\frac{1}{8}(t_1x_1+t_2x_2),\quad
  \beta_T = \frac{5}{24}(T+U).
\end{align}\end{subequations}

\section{Results and Discussion}\label{sec:result}

The fitting protocol of SAMi-T is similar to that of SAMi \cite{Roca-Maza2012}, with further constraint on the tensor term from the pseudodata of spin-orbit splitting evolution in neutron-proton drops by RBHF calculations using the Bonn A interaction \cite{Machleidt1989}.
The detail of the fitted data and fitted results has been reported in Ref.~\cite{Shen2018c}.
The new functional SAMi-T can give a good description of ground state properties such as binding energy and charge radii (with an accuracy of $1\%$ for several selected doubly magic nuclei), excited state properties such as GMR, GDR, GTR, and SDR \cite{Shen2018c}.
In this contribution we focus on the spin-orbit splittings of finite nuclei given by SAMi-T.

\begin{figure}
\centering
\includegraphics[width=8cm]{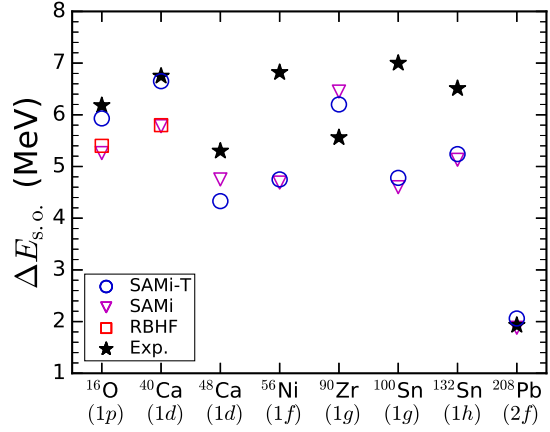}
\caption{(Color online) Proton spin-orbit splittings of finite nuclei calculated by SAMi-T, in comparison with experimental data, results of SAMi functional, and RBHF theory using the Bonn A interaction.}
\label{fig:so-fn}
\end{figure}

In Fig.~\ref{fig:so-fn}, the proton spin-orbit splittings in several finite nuclei calculated by SAMi-T are shown, in comparison with experimental data, with the results of SAMi, and of RBHF theory using the Bonn A interaction.
For $^{90}$Zr and $^{208}$Pb, SAMi-T and SAMi give similar results, while for the
doubly-closed-shell nuclei $^{16}$O and $^{40}$Ca, the spin-orbit splittings given by SAMi-T agree well   with experimental data but those given by SAMi are slightly smaller.
This is understandable as both SAMi-T and SAMi have fitted to the SO splittings of $^{90}$Zr and $^{208}$Pb, but tensor terms ($\alpha$ and $\beta$) been reduce the SO splittings in these two spin-unsaturated nuclei.
Therefore, to reproduce the same data in these two nuclei SAMi-T need larger SO terms ($W_0$ and $W_0'$).
For spin-saturated nuclei like $^{16}$O and $^{40}$Ca, the tensor terms give no contribution and with larger SO terms SAMi-T will give larger SO splittings.

The spin-orbit splittings given by RBHF are also smaller than the data.
Since the relative change of SO splittings as the particle numbers contains the information of tensor force, when we fit the spin-orbit splittings of neutron-proton drops given by RBHF theory, we only fit the relative change instead of the absolute value.
As the \textit{ab initio} RBHF calculation starts with the bare nuclear force which is fitted to nucleon-nucleon scattering, there is no free parameter and the information of tensor force by RBHF shall be reliable in this sense.

For nuclei $^{48}$Ca, $^{56}$Ni, $^{100}$Sn, and $^{132}$Sn, the spin-orbit splittings given by SAMi-T and SAMi are both smaller than the data, while for $^{90}$Zr they are both larger.

\section{Summary}\label{sec:sum}

In this work we developed a new Skyrme functional SAMi-T with tensor term guided by \emph{ab initio} relativistic Brueckner-Hartree-Fock calculations.
Using the spin-orbit splittings of neutron-proton drops calculated by RBHF, which does not contain beyond-mean-field effect such as particle-vibration coupling, the tensor terms of SAMi-T have been well constrained.
Besides that, SAMi-T follows the fitting protocol of the successful SAMi functional, and it can well describe the ground state properties (such as binding energy, charge radii, and spin-orbit splittings) and excited state properties (such as GMR, GDR, GTR, and SDR) of finite nuclei \cite{Shen2018c}.

We have compared the spin-orbit splittings of several nuclei given by SAMi-T and by SAMi.
It was found for spin-saturated nuclei such as $^{16}$O and $^{40}$Ca the SO splittings by SAMi are slightly smaller than the experimental data and SAMi-T improves in this case.

\vspace{2em}
Funding from the European Union's Horizon 2020 research and innovation programme under Grant agreement No. 654002 is acknowledged.


\end{document}